\documentclass[amsmath, amssymb, superscriptaddress, reprint, twocolumn, colorlinks=true, citecolor=blue, linkcolor=blue, urlcolor=blue]{revtex4-1}
\usepackage{natbib}
\usepackage[utf8]{inputenc}
\usepackage[T1]{fontenc}
\usepackage{graphicx}
\usepackage{grffile}
\usepackage{longtable}
\usepackage{wrapfig}
\usepackage{rotating}
\usepackage[normalem]{ulem}
\usepackage{amsmath}
\usepackage{textcomp}
\usepackage{amssymb}
\usepackage{capt-of}
\usepackage{hyperref}
\usepackage{mathptmx}
\usepackage{upgreek}
\usepackage{float}
\usepackage{bm}
\usepackage{setspace}
\usepackage{booktabs}
\usepackage{subfigure}
\usepackage{url}
\usepackage{xcolor}
\usepackage{multirow}
\usepackage{array}
\usepackage{tabularx}
\usepackage{nicematrix}
\usepackage{blkarray}
\usepackage{mathptmx}
\usepackage{mathrsfs}
\usepackage{xcolor}
\makeatletter
\newcommand{\coloredcite}[1]{{\color{ccr}\cite{#1}}}
\newcommand{\coloredeqref}[1]{{\color{ccr}\eqref{#1}}}
\makeatother
\makeatletter
\renewcommand{\maketag@@@}[1]{\hbox{\m@th\normalsize\normalfont#1}}
\makeatother

\definecolor{ccr}{RGB}{45,47,145} 

\hypersetup{colorlinks=true, citecolor=ccr, urlcolor=ccr, linkcolor=ccr}

\DeclareMathAlphabet{\mathcal}{OMS}{cmsy}{m}{n}

\date{\today}

\begin{document}

\title{Generalized Similarity Theory for Plasmas}

%\title{Generalized Theory of Plasma Similarity}

\author{Yangyang Fu}
\email{fuyangyang@tsinghua.edu.cn}
\affiliation{Department of Electrical Engineering, Tsinghua University, Beijing 100084, China}
\affiliation{State Key Laboratory of Power System Operation and Control, Department of Electrical Engineering, Tsinghua University, Beijing 100084, China}
%\affiliation{Sichuan Energy Internet Research Institute, Tsinghua University, Sichuan 610213, China}

% \author{Chubin Lin}
% \affiliation{Department of Electrical Engineering, Tsinghua University, Beijing 100084, China}

% \author{Jiandong Chen}
% \affiliation{Department of Electrical Engineering, Tsinghua University, Beijing 100084, China}

% \author{Hanyang Li}
% \affiliation{Department of Electrical Engineering, Tsinghua University, Beijing 100084, China}

% \author{Zhen Wang}
% \affiliation{Department of Electrical Engineering, Tsinghua University, Beijing 100084, China}

% \affiliation{State Key Laboratory of Power System Operation and Control, Department of Electrical Engineering, Tsinghua University, Beijing 100084, China}
% \affiliation{Sichuan Energy Internet Research Institute, Tsinghua University, Sichuan 610213, China}

\begin{abstract}
A generalized theory of plasma similarity is established based on the scaling of the Boltzmann equation coupled with the full set of Maxwell's equations.
The predicted similarity scalings are demonstrated through first-principles particle-in-cell and fluid simulations across diverse operational regimes, including collisionless plasmas with two-stream instabilities, pure electron diodes spanning the classical to relativistic regimes, electromagnetic-wave-driven plasmas, and discharge plasmas ranging from low to high ionization regimes.
The generalized theory reveals the intrinsic scale-invariant nature of plasmas under specified conditions, rooted in the symmetry and scaling transformations of the governing equations.
\end{abstract}

\maketitle
\textit{Introduction}---Similarity methods have been developed in diverse physical systems \coloredcite{sedov2018similarity,Bluman,Bridgman}, revealing scale-invariant physical laws under upscaled or downscaled conditions \coloredcite{ryutov2018scaling,sauppe2020demonstration}, thereby providing fundamental insights for predicting, mapping, and correlating the complex dynamics of parameters across spatiotemporal scales.
Similarity laws have proven applicable to various plasma systems, including glow discharges \coloredcite{Engel-597,MichaelKhodorkovskii-600,Gudmundsson_2017}, streamers \coloredcite{pasko2007red,LiuPasko}, pulsed discharges \coloredcite{Mesyats-2006}, multipactor discharges \cite{woo1967a,lau2006scaling,peng100}, microdischarges \coloredcite{Janasek-nature,loveless2016scaling}, gas discharge lasers \cite{gordon1963similarity}, fusion plasmas \coloredcite{lacina1971application,luce2008application}, and high-frequency plasmas \coloredcite{Lisovskiy_2008EPL,ferreira1988the}.
According to classical similarity laws \coloredcite{fu2018similarities_tps}, the product of gas pressure and gap dimension \(pd\) (reduced length) and the ratio of electric field to gas pressure \(E/p\) (reduced electric field) are identified as the scale-invariant combined parameters. These two combined parameters govern fundamental behaviors of discharge plasmas, including the breakdown voltage (e.g., Paschen's law \cite{paschen1889onENG}) and the electron-impact neutral ionization rate coefficient (e.g., Townsend theory \cite{Townsend-596}). These descriptions, however, are contingent upon the local-field or local-energy approximations, which are not always valid \coloredcite{Fu-RMPP2023}.
Previous studies \coloredcite{Yd-psst,FuZheng-pop2020,fu-prapplied,Yang-2023,YANGDONGapl}, via particle-in-cell (PIC) simulations, have demonstrated the similarity law for low-pressure radio-frequency (rf) discharges, where electron kinetic behaviors are highly nonlocal.
Subsequently, the similarity relation for the electron excitation rate in capacitive rf plasmas has been experimentally validated via phase-resolved optical spectroscopy \coloredcite{PRL}.
Recently, similarity laws have been applied to pure electron systems (e.g., electron-beam-driven vacuum diodes \coloredcite{lcb-edl,lcb-pre}), demonstrating theoretically rigorous scale-invariant relations under designed conditions where complex plasma chemistry is not a concern.
To date, however, similarity laws remain limited to stabilized charge transport systems or low-temperature plasmas in weakly ionized regimes, without explicitly accounting for the roles of electromagnetic effects in high-frequency plasmas, gas heating in thermal plasmas, or relativistic electron beams in diodes.
Consequently, a generalized similarity theory for plasmas remains to be developed to uncover their fundamental scale-invariant nature across a broad range of conditions.

In this Letter, a generalized similarity theory is established for plasmas from the Boltzmann equation and the full set of Maxwell's equations.
The similarity scalings are demonstrated for diverse plasma systems, including collisionless plasmas with two-stream instabilities, pure electron diodes from non-relativistic to relativistic regimes, capacitive rf plasmas with strong electromagnetic effects, and microdischarge plasmas from low to high ionization degrees.
This work offers a comprehensive perspective that extends the similarity theory and reveals the fundamental scale-invariant symmetry underlying a wide range of plasma sources.

\textit{Similarity transformation}---The plasma behaviors are governed by the Boltzmann equation, which describes the distribution function \(f_j=f_j(\textbf{r},\textbf{v},t)\) for species \(j\) and is given by
\begin{equation}
\frac{\partial f_j}{\partial t} + \mathbf{v} \cdot \frac{\partial f_j}{\partial \mathbf{r}} + \frac{q_j}{m_j}(\mathbf{E}+\mathbf{v}\times \mathbf{B}) \cdot \frac{\partial f_j}{\partial \mathbf{v}}  = \left(\frac{\partial f_j}{\partial t}\right)_{\text{coll}},
\label{eq:boltzmann0}
\end{equation}
where \(\textbf{r}\) is the position, \(\textbf{v}\) is the velocity, \(q_j\) is the species charge, and \(m_j\) is the species mass; \(\textbf{E}\) and \(\textbf{B}\) are the electric and magnetic fields; \((\partial f_j/\partial t)_\text{coll}\) is the collisional term, including charge-neutral and charge-charge collisions \coloredcite{lieberman2005principles}.
The electromagnetic fields in the Boltzmann equation are described by Maxwell's equations, which are expressed as
\begin{subequations}
\label{eq:Maxwell}
\begin{align}
 \nabla_{\mathbf{r}}\cdot \mathbf{E}
  & =\frac{1}{\varepsilon_0}\sum_{j}q_j\int f_{j} \mathrm{d}\mathbf{v},\label{maxwelleq1}\\
 \nabla_{\mathbf{r}}\cdot\mathbf{B}
&  = 0 ,\\
 \nabla_{\mathbf{r}}\times\mathbf{E}
 & = \frac{\partial \mathbf{B}}{\partial t},\\
 \nabla_{\mathbf{r}}\times \mathbf{B} 
 & ={\mu_0}\sum_{j }^{} q_j \int  f_j\mathbf{v}\text{d}\mathbf{v} +{\mu_0}\varepsilon_0\frac{\partial \mathbf{E}}{\partial t},
 \label{maxwelleq}
\end{align}
\end{subequations}
where \(j\in\{e,i\}\) denotes the electron and ion species, \(\varepsilon_0\) is the vacuum permittivity, and \(\mu_0\) is the vacuum permeability.

The similarity scaling is first derived for the Vlasov-Poisson system, corresponding to collisionless plasmas in the electrostatic regime. The Vlasov equation is expressed as
\begin{equation}
\label{eq:VP1}
 \frac{\partial f_j}{\partial t} + \mathbf{v} \cdot \frac{\partial f_j}{\partial \mathbf{r}} + \frac{q_j}{m_j}(\mathbf{E}+\mathbf{v}\times \mathbf{B}) \cdot \frac{\partial f_j}{\partial \mathbf{v}}  = 0.
\end{equation}

Scaling the Vlasov equation Eq.~\eqref{eq:VP1} by a factor of \(k^3\) yields
\begin{equation}
\label{eq:VP3}
\begin{aligned}
\frac{\partial (k^{-2}f_j)}{\partial (kt)} + \mathbf{v} \cdot \frac{\partial (k^{-2}f_j)}{\partial (k\mathbf{r})} & + \\ \frac{q_j}{m_j} [ k^{-1}\mathbf{E}& + \mathbf{v}\times (k^{-1}\mathbf{B})] \cdot \frac{\partial (k^{-2}f_j)}{\partial \mathbf{v}}  = 0.
\end{aligned}
\end{equation}

The Poisson equation Eq.~\eqref{maxwelleq1} scaled by a factor of \(k^2\) can be formulated as
\begin{equation}
\label{eq:VP4}
\nabla_{k\mathbf{r}}\cdot (k^{-1}\mathbf{E})
  =\frac{1}{\varepsilon_0}\sum_{j}q_j\int (k^{-2}f_{j}) \mathrm{d}\mathbf{v}.
\end{equation}

Comparing Eqs.~\eqref{maxwelleq1} and \eqref{eq:VP1}--\eqref{eq:VP4}, one finds that the solutions of the corresponding Vlasov–Poisson systems are scale-invariant provided that the variables are scaled as \(\{k^{-2}f_j, kt, k\mathbf{r}, k^{-1}\mathbf{E}, k^{-1}\mathbf{B}\}\) from one system to the other, while \(\mathbf{v}\), \(q_{j}\), and \(m_{j}\) remain unchanged.
Thus, for a prototype and a downscaled system, the parameters transform as \(\mathbf{r}_1=k\mathbf{r}_k\), \({t}_1=kt_k\), \(\mathbf{E}_1=k^{-1}\mathbf{E}_k\), \(\mathbf{B}_1=k^{-1}\mathbf{B}_k\), \(\mathbf{v}_1=\mathbf{v}_k\), and \(f_{j,1}=k^{-2}f_{j,k}\), with \(k\) being the scaling factor. 

The similarity transformations for the physical parameter \(G(\textbf{r},t)\) in the prototype (subscript 1) and the scaled system (subscript $k$) can be mathematically generalized using
\begin{equation}
G(\mathbf{r}_1, t_1) = k^{\alpha[G]} G(\mathbf{r}_k, t_k),
\label{eq-G1Gk}
\end{equation}
where \(\alpha[\mathbf{r}]=\alpha[{t}] = 1\), \(\alpha[\mathbf{v}] = 0\), \(\alpha[\mathbf{E}] = \alpha[\mathbf{B}] = -1\), and \(\alpha[f_{j}] = -2\), yielding similarity scaling for Vlasov–Poisson systems in the electrostatic regime.

\textit{Electromagnetic effects}---The similarity scaling is not limited to the electrostatic fields. When electromagnetic effects in plasmas become significant, the fields are described by the full set of Maxwell's equations, rather than Poisson's equation alone. The scaling of Maxwell's equations, scaled by a factor of \(k^2\), can be reformulated as
\begin{subequations}
\label{eq:Maxwell2}
\begin{align}
 \nabla_{k\mathbf{r}}\cdot (k^{-1}\mathbf{E}) 
  & =\frac{1}{\varepsilon_0}\sum_{j}q_j\int k^{-2}f_{j} \mathrm{d}\mathbf{v},\\
 \nabla_{k\mathbf{r}}\cdot(k^{-1}\mathbf{B}) 
&  = 0 ,\\
 \nabla_{k\mathbf{r}}\times(k^{-1}\mathbf{E}) 
 & = \frac{\partial (k^{-1}\mathbf{B})}{\partial (kt)},\\
 \nabla_{k\mathbf{r}}\times (k^{-1}\mathbf{B}) 
 & ={\mu_0}\sum_{j }^{} q_j \int  k^{-2} f_j\mathbf{v}\text{d}\mathbf{v}+{\mu_0}\varepsilon_0\frac{\partial (k^{-1}\mathbf{E})}{\partial (kt)}.
\end{align}
\end{subequations}
The scaled variables \(\{k^{-2}f_j, kt, k\mathbf{r}, k^{-1}\mathbf{E}, k^{-1}\mathbf{B}\}\) are consistent with the scaling from the Boltzmann equation and thus confirm the similarity scaling in the electromagnetic regime.

\textit{Relativistic effects}---In the pure charge systems (e.g., electron diodes \coloredcite{lcb-edl,lcb-pre}), relativistic effects must be considered when the particle velocity approaches the speed of light \(c\).
To account for relativistic effects, the Boltzmann equation should be cast into its relativistic form \cite{Akama_1970}, given by
\begin{equation}
\frac{\mathcal{E}}{c^2}\frac{\partial f_j}{\partial t} + \mathbf{p} \cdot \frac{\partial f_j}{\partial \mathbf{r}} + q_j(\mathbf{E}+\mathbf{v}\times \mathbf{B}) \cdot \frac{\partial f_j}{\partial \mathbf{p}}  = 0,
\label{eq:relativistic}
\end{equation}
where \(f_j = f(\textbf{r}, \textbf{p}, t)\) is the distribution function of species \(j\), \(\mathcal{E}=\sqrt{\mathbf{p}^2c^2+m_j^2c^4}\), \(\textbf{p}=\gamma m_j\textbf{v}\) is the momentum of species, and \(\gamma=1/\sqrt{1-\mathbf{v}^2/c^2}\) is the Lorentz factor.

Dividing Eq.~\coloredeqref{eq:relativistic} by a factor of \(k^3\) yields
\begin{equation}
\begin{aligned}
    \frac{\mathcal{E}} {c^2}\frac{\partial (k^{-2}f_j)}{\partial (kt)} &+ \mathbf{p} \cdot \frac{\partial (k^{-2}f_j)}{\partial (k\mathbf{r})}  + \\ q_j[k^{-1}\mathbf{E}&+\mathbf{v}\times (k^{-1}\mathbf{B})] \cdot \frac{\partial (k^{-2}f_j)}{\partial \mathbf{p}}  = 0.
    \label{eq:relativistic2}
\end{aligned}
\end{equation}
The mathematical solutions of Eqs.~\eqref{eq:relativistic}--\eqref{eq:relativistic2} are scale-invariant with scaled physical variables \(\{k^{-2}f_e\), \(k\mathbf{r}\), \(kt\), \(k^{-1}\mathbf{E}\), \(k^{-1}\mathbf{B}\}\) in compared systems. The full set of Maxwell's equations is inherently compatible with relativistic effects \coloredcite{Cario}. More fundamentally, the Lorentz factor \(\gamma\) is scale-invariant under similarity transformations, thereby preserving similarity scalings in the relativistic regime.

The similarity transformation for plasmas roots in the symmetry nature of Boltzmann-Maxwell equations, indicating that plasma characteristics can be invariant across physical scales \coloredcite{Fu-RMPP2023}. The similarity laws hold when collisional effects are considered, spanning charge-neutral to charge-charge-dominated collisional regimes (see Appendix \hyperlink{appendixA}{A}). In Appendix \hyperlink{appendixA}{B}, the similarity scalings are mathematically derived using dimensional analysis.
In the following, the universality of the similarity framework is demonstrated across a diverse range of plasma systems, including plasma two-stream instabilities, pure-electron diodes, capacitive rf plasmas, and microdischarges spanning the glow-to-arc regimes.

\begin{figure}[htbp]
\hypertarget{fig1}{}
\centering
\includegraphics[clip, width=1\linewidth]{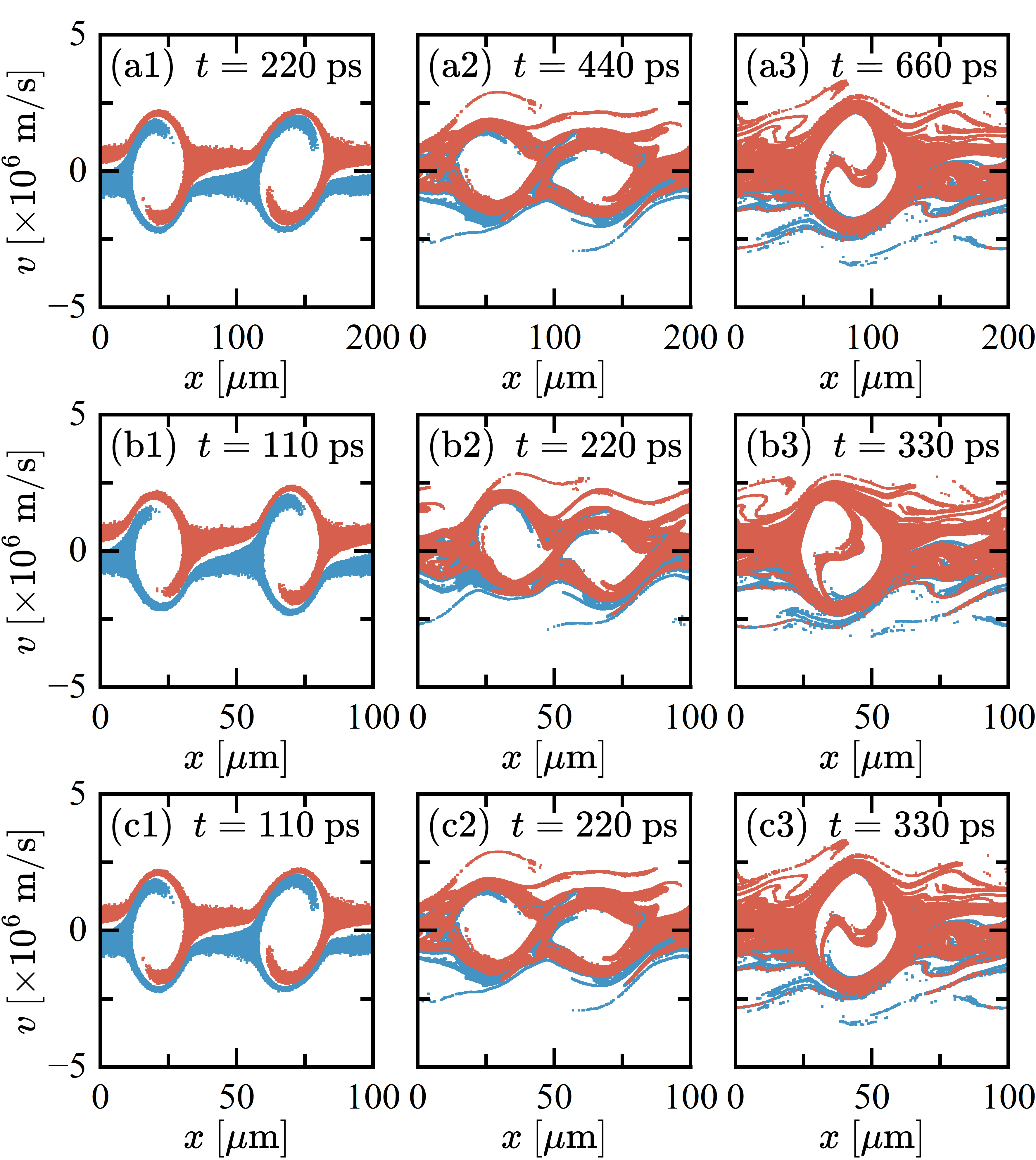}
\caption{\label{fig:Fig1} Dynamical similarity in plasma two-stream instabilities. Velocity phase space distributions of the two-electron-stream instability for the base case [(a1)--(a3)] and for scaled conditions without [(b1)--(b3)] and with [(c1)--(c2)] identical initial states, respectively.}
\end{figure}

\textit{Plasma two-stream instabilities}---The dynamical similarities in plasma two-stream instabilities are shown in Fig.~\hyperlink{fig1}{1}.
The velocity phase-space distributions of the electron two-stream instability under scaled conditions are obtained for two-electron beams via PIC simulations. The time and gap dimensions are scaled by a factor of two.
For the prototype case (\(k=1\)) [Figs.~\hyperlink{fig1}{1(a1)}--\hyperlink{fig1}{1(a3)}], the velocity phase-space distributions at varied time explicitly demonstrate the presence of electron holes, which are induced by the two counter-streaming electron beams.
In the proportionally scaled cases (\(k=2\)) [Figs.~\hyperlink{fig1}{1(b1)}--\hyperlink{fig1}{1(b3)}], the electron phase-space distributions at corresponding scaled time steps are rather the same as those in the prototype cases. Since the dynamical evolutions of two-stream instability are significantly affected by the initial particle states of the statistical systems, the profiles of the phase space are not exactly the same in the compared systems. If the initial states of the compared systems are kept identical, the plasma two-stream instabilities are shown to be exactly the same [Figs.~\hyperlink{fig1}{1(c1)}--\hyperlink{fig1}{1(c3)}].
Here, the PIC results confirm that the similarity scaling holds rigorously for plasmas under non-steady-state conditions.
Theoretically, the scale-invariant two-stream instability in plasmas can be understood from the similarity derivation using Vlasov-Poisson equations.

\textit{Non-relativistic to relativistic diodes}---Relativistic effects are essential for applications of intense electron-beam diodes, accelerators, ultrafast electron microscope, and free-electron lasers, where the kinetic energy of charged particles becomes comparable to the rest-mass energy \coloredcite{Benford}.
Here, the similarity method is developed from the non-relativistic regime to the relativistic regime, and the similarity scaling is demonstrated for the electron beam-driven diode.
When an intense electron beam is injected from the cathode and transported toward the anode in a pure electron diode, the electron velocity can be tuned by the applied electric field from the non-relativistic to the relativistic regime.
Figure~\hyperlink{fig2}{2} shows the scale-invariant nature of the electron velocity phase space and the time-dependent cathode surface electric field in the compared systems via PIC simulations.
The distributions of electron velocity phase space \((v-kx)\) with scaled spatial position overlap in the compared cases, where \(k=1\) indicates the prototype case and \(k=2\) denotes the scaled case.
The electron velocity \(v\) is far less than the speed of light \(c\) in the non-relativistic regime [Fig.~\hyperlink{fig2}{2(a)}], while it approaches the speed of light, \(v\sim c\), in the relativistic regime [Fig.~\hyperlink{fig2}{2(b)}].
The oscillation is due to space-charge effects \coloredcite{lcb-pre}, and becomes within sub-picoseconds in the relativistic case.
The scaled oscillating cathode electric fields, shown by \((k^{-1}E_\text{s}-kt)\), coincide in both non-relativistic and relativistic regimes [Figs.~\hyperlink{fig2}{2(c)} and \hyperlink{fig2}{2(d)}].
The results demonstrate the applicability of the similarity method from non-relativistic to relativistic regimes.

\begin{figure}[htbp]
\hypertarget{fig2}{}
\centering
\includegraphics[clip, width=\linewidth]{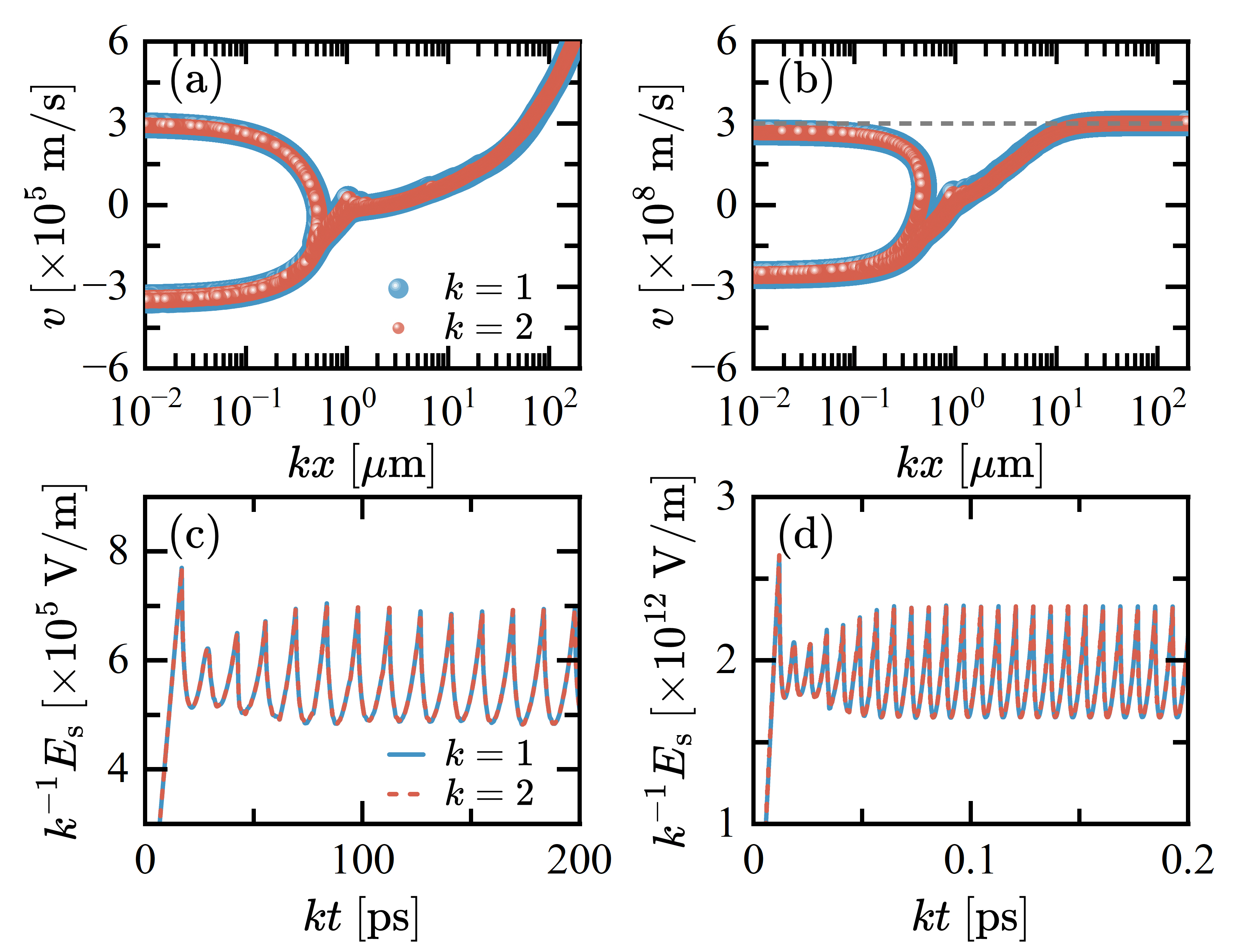}
\caption{\label{fig:Fig2}
Similarity scaling for electron-beam-driven diodes. (a)--(b) Velocity phase space distributions for prototype (\(k=1\)) and downscaled (\(k=2\)) diodes versus scaled position in (a) non-relativistic and (b) relativistic regimes. (c)--(d) Scaled cathode electric field versus scaled time in (c) non-relativistic and (d) relativistic regimes.
}
\end{figure}

\textit{Standing wave effects in rf plasmas}---Similarity laws have been demonstrated for capacitive rf plasma in the electrostatic regime. However, for plasmas operating at very high frequencies and large areas, electromagnetic effects (e.g., standing-wave effects \coloredcite{2019prlzhaokai}) can become pronounced. In such cases, the field equations should be described by Maxwell's equations, rather than Poisson's equation alone \coloredcite{Lieberman2002PSST}.
%The similarity laws are consistently established from the scaling of the Boltzmann equation and Maxwell's equations.
Here, a two-dimensional electromagnetic PIC code \coloredcite{Eremin2025,Eremin_2023,li2026similaritytheoryscalingnetworks} is employed to demonstrate the scale-invariant nature of rf plasmas with pronounced standing-wave effects.
In the simulations, three electron-neutral collisions (elastic, excitation, and ionization) and two ion-neutral collisions (isotropic and backward scattering) are considered. As shown in Fig.~\hyperlink{fig3}{3}, in two geometrically similar gaps, \([p, z, r, f]=[40\ \text{mTorr}, 6\ \text{cm}, 38\ \text{cm}, 106 \ \text{MHz}]\) for the base case and \([80\ \text{mTorr}, 3\ \text{cm}, 19\ \text{cm}, 212\ \text{MHz}]\) for the down-scaled case (with scaling factor \(k = 2\)), the electron density distributions remain identical, exhibiting the same perturbations.
The plasma nonuniformity, featuring multiple radial density peaks in the bulk, is due to standing wave effects when the reactor size becomes comparable to the excitation wavelength of electromagnetic waves.
The ratio of the plasma density obeys the similarity scaling \(n_{e,k}/n_{e,1} = k^2\) with \(k=2\), e.g., the maximum densities are \(3.0 \times 10^{16} \) \(\text{m}^{-3}\) and \(1.2 \times 10^{17}\) \(\text{m}^{-3}\), respectively.
The results unambiguously confirm that similarity holds for plasmas with strong electromagnetic effects across varying parameter scales, extending the similarity framework from the electrostatic to electromagnetic regimes.

\begin{figure}[htbp]
\hypertarget{fig3}{}
\centering
\includegraphics[clip, width=\linewidth]{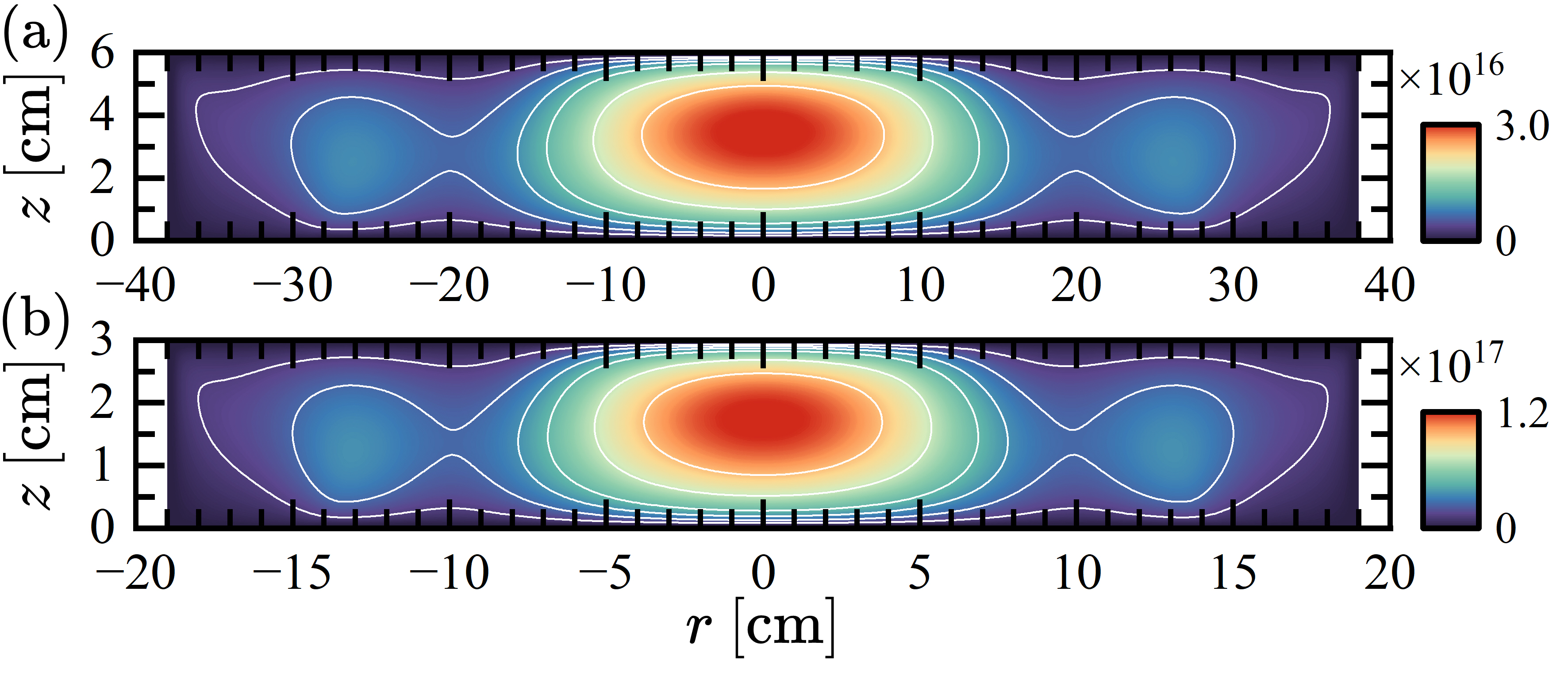}
\caption{\label{fig:Fig3} 
Spatial distributions of electron density in capacitive rf plasmas for (a) the base case and (b) the downscaled case, both with strong electromagnetic (standing wave) effects.
}
\end{figure}

\begin{figure}[htbp]
\hypertarget{fig4}{}
\centering
\includegraphics[clip, width=\linewidth]{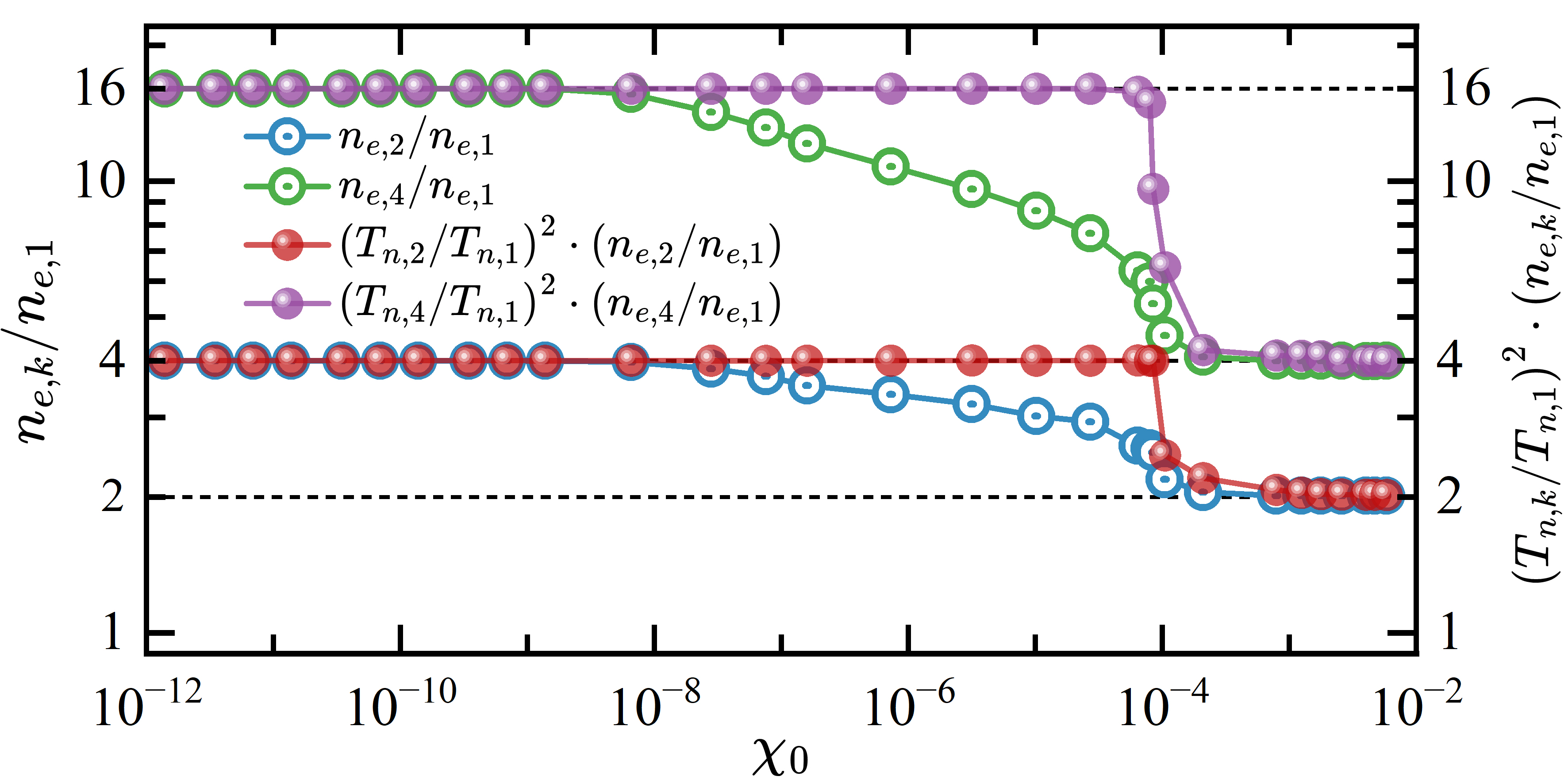}
\caption{\label{fig:Fig4}
Similarity scaling of the electron density ratio versus the ionization degree in high-pressure microdischarges.
The modified similarity scaling accounts for gas heating and extends the applicability across regimes from low to high ionization degrees.
}
\end{figure}

\textit{Low to high ionization degree regimes}---The similarity laws are mostly developed for low-temperature plasmas in weakly ionized regimes \coloredcite{Fu-RMPP2023}.
However, for high-pressure microdischarges with strong ionization, the effects of gas heating and consumption of neutral gases can be important, and the conventional similarity scaling could be questionable \coloredcite{Fu_2019}.
According to the theoretical derivation (Appendix \hyperlink{appendixA}{A}), charge-charge interactions can dominate collisions when the background gas is almost fully ionized.
Here, the similarity scaling is determined for high-pressure microdischarges via a unified fluid model \coloredcite{wang_2024,baeva2019unified,baeva2020unified}, which incorporates heat transfer between the plasma and the cathode and captures the transition across the Townsend–glow-arc regimes [see Ref.~\coloredcite{wang_2024} for more details].
The microdischarges in the systems being compared are simulated under similarity conditions.
Figure~\hyperlink{fig4}{4} presents the ratio of electron density \(n_{e,k}/n_{e,1}\) versus the ionization degree \(\chi_0\) between the base case \((k=1)\) and downscaled systems \((k=2\) and \(k=4)\) for discharges from low to high ionization degree regimes.
The electron density scales as \(n_{e,k}/n_{e,1} = k^2\) in weakly ionized plasmas and as \(n_{e,k}/n_{e,1} = k\) in high ionization degree regimes, showing significant deviations in the transition.
Since gas heating and complex chemical reactions cause similarity violations during the transition, a modified dimensionless scaling, \({(T_{n,k}/T_{n,1})}^2\cdot(n_{e,k}/n_{e,1})\) versus \(\chi_0\), incorporating the effect of neutral gas temperature, is proposed to generalize similarity scaling from weak to strong ionization regimes.
The modified scaling holds for \(\chi_0 \rightarrow 0\) (electron-impact neutral ionization dominates the generation of charged particles) and for \(\chi_0 \rightarrow 1\) (electron-electron interactions dominate), respectively.
While less mathematically rigorous, the modified similarity scaling significantly reduces the violation during the transition, thereby effectively extending its applicability from low to high ionization regimes.

\textit{Conclusions}---In conclusion, this Letter generalized and demonstrated the similarity theory for plasmas across diverse operational regimes, including two-stream plasma instability and the transitions from non-relativistic to relativistic, electrostatic to electromagnetic, and weakly to highly ionized regimes.
The similarity scaling derived from the Boltzmann equation and Maxwell's equations mathematically unravels the invariance and homogeneity of the plasma governing equations.
The generalized similarity theory uncovers the inherent scale-invariant nature of plasmas under designed conditions, substantially extending the applicability of the similarity framework across a wide range of plasma sources.

\textit{Acknowledgments}---The author acknowledges the financial support from the Fundamental and Interdisciplinary Disciplines Breakthrough Plan of the Ministry of Education of China (Grant No.~JYB2025XDXM312).

%\vspace{1em}
\textit{Data availability}---The data that support the findings of this study are available from the contact author upon reasonable request.

% \Large
\bibliography{references}

\appendix
\onecolumngrid
\begin{center}
\vspace{1em}
\large\textbf{End Matter}
\end{center}
\twocolumngrid

%\begin{center}

\hypertarget{appendixA}{}\textit{Appendix A: Collisional effects}---The similarity scaling for plasmas can be derived from the Boltzmann equation, with consideration of the scaling of the collisional term. The Boltzmann equation for electrons is expressed as
\begin{equation}
\frac{\partial f_e}{\partial t} + \mathbf{v} \cdot \frac{\partial f_e}{\partial \mathbf{r}} + \frac{e}{m_e}(\mathbf{E}+\mathbf{v}\times \mathbf{B}) \cdot \frac{\partial f_e}{\partial \mathbf{v}}  = \left(\frac{\partial f_e}{\partial t}\right)_{\text{coll}},
\label{eq:boltzmann0}
\end{equation}
where \(e\) is the elementary charge, and \(m_e\) is the electron mass. The collisional term \((\partial f_e/\partial t)_\text{coll}\), composed of electron-neutral and electron-electron collisions \coloredcite{lieberman2005principles}, is given by
\begin{equation}
\left(\frac{\partial f_{e}}{\partial t}\right)_{\text{coll}}= C_\text{en}(f_{e},f_{n})+C_\text{ee}(f_{e},f_{e}),
\label{eq:collision0}
\end{equation}
where \(f_{e}\) and \(f_{n}\) are the distribution functions of electrons and neutrals. The electron-neutral and electron-electron collisional terms are expressed as
\begin{subequations}
\begin{align}
C_\text{en}(f_{e},f_{n})&=\iint\left(f_{e2}f_{n2}-f_{e1}f_{n1}\right)\mathbf{v}_\text{en}\sigma_\text{en}\text{d}\Omega\text{d}\mathbf{v},\label{eq:collision0}\\
C_\text{ee}(f_{e},f_{e})&=\iint\left(f_{e2}f_{e2}-f_{e1}f_{e1}\right)\mathbf{v}_\text{ee}\sigma_\text{ee}\text{d}\Omega\text{d}\mathbf{v},
\label{eq:collision1}
\end{align}
\end{subequations}
where \((f_{e1}\), \(f_{n1})\) and \((f_{e2}\), \(f_{n2})\) denote the distribution functions of electrons and neutrals before and after collisions, respectively; \(\mathbf{v}_\text{en}\) and \(\mathbf{v}_\text{ee}\) are the relative velocities, \(\sigma_\text{en}\) and \(\sigma_\text{ee}\) are the cross sections, and \(\text{d}\Omega\) is the solid angle element.

\textit{For weakly ionized plasmas}---For the plasmas in weakly ionized regimes, electron-neutral interactions dominate the collision. Dividing the collisional term Eq.~\eqref{eq:collision0} by \(k^3\) yields
\begin{equation}
\label{weaksl}
    \frac{1}{k^3}C_\text{en}(f_{e},f_{n})=\iint\left(\frac{f_{e2}}{k^{2}}\frac{f_{n2}}{k}-\frac{f_{e1}}{k^{2}}\frac{f_{n1}}{k}\right)\mathbf{v}_\text{en}\sigma_\text{en}\text{d}\Omega\text{d}\mathbf{v},
\end{equation}
where the scaled invariance of the electron distribution function \(k^{-2}f_e\) holds with \(k^{-1}f_n\) being scale-invariant, which can be maintained by scaling the neutral gas pressure, i.e., keeping \(k^{-1}p\) constant in compared systems. For discharge plasmas with the gap dimension \(d\), gas pressure \(p\), and driving voltage frequency \(f\) appropriately scaled, i.e., keeping \(pd\) and \(fd\) (or \(f/p\)) constant, plasma parameters obey the similarity scaling, explicitly demonstrating an intrinsic scale-invariant nature in compared systems.

\textit{For strongly ionized plasmas}---In the plasmas in highly ionized regimes, the charge-charge interaction becomes dominant in the collisional term, whereas the electron-neutral collisions have minor effects. Here, from Eq.~\eqref{eq:collision1} the collisional term with electron-electron collisions being dominant is scaled as follows
\begin{equation}
    \frac{1}{k^2}C_\text{ee}(f_{e},f_{e})=\iint\left(\frac{f_{e2}}{k^{}}\frac{f_{e2}}{k}-\frac{f_{e1}}{k^{}}\frac{f_{e1}}{k}\right)\mathbf{v}_\text{ee}\sigma_\text{ee}\text{d}\Omega\text{d}\mathbf{v},
\end{equation}
where \(k^{-1}f_e\) becomes scale-invariant in compared systems.
Note that in high-density quasineutral plasmas, the electric field \(\mathbf{E}\), rather than being solved by Poisson's equation, can be obtained from the ambipolar equation through
\begin{equation}
\mathbf{E}=-\frac{T_e}{n_e}\frac{\partial n_e}{\partial \mathbf{r}},
\label{eq-bipolar}
\end{equation}
where \(T_e\) is the electron temperature and \(n_e\) is the electron density. Therefore, in the highly ionized plasma regime, \(\alpha[n_e]=\alpha[f_e] =-1\) since \(n_e=\int f_e(\mathbf{v})\text{d}\mathbf{v}\) and \(\alpha[\mathbf{v}]=0\). The similarity scaling for the electric field is expressed as
\begin{equation}
k^{-1}\textbf{E}=-\frac{T_e}{k^{-1}n_e}\frac{\partial (k^{-1}n_e)}{\partial (k\mathbf{r})},
\label{eq-bipolar1}
\end{equation}
where \(\alpha[\mathbf{E}]=-1\) and \(\alpha[T_e]=0\) are preserved and consistent with the similarity scaling in weakly ionized plasmas.
The ambipolar electric field can exist in both weak and strong ionization regimes. From Eq.~\eqref{weaksl}, \(\alpha[n_e]=\alpha[f_e] =-2\) seems to betray the similarity of the ambipolar mechanism in weakly ionized plasmas. However, the ambipolar field can be scaled to satisfy the similarity law as follows
\begin{equation}
k^{-1}\textbf{E}=-\frac{T_e}{k^{-2}n_e}\frac{\partial (k^{-2}n_e)}{\partial (k\mathbf{r})}.
\label{eq-bipolar2}
\end{equation}
Equations~\eqref{eq-bipolar1}--\eqref{eq-bipolar2} confirm that ambipolar mechanisms obey similarity scaling across weak-to-strong ionization regimes.

\hypertarget{appendixA}{}\textit{Appendix B: Dimensional analysis}---Consider two plasma systems (systems I and II) that satisfy the similarity requirements of conditions; the physical quantity $G$ in systems I and II can be transformed by
\begin{equation}
    G_{\text{I}} = a_G G_{\text{II}},
    \label{eq15}
\end{equation} 
where $a_G$ is a dimensionless coefficient.
More specifically, for the eight related quantities in the Boltzmann equation, one can have \(f_{\text{I}} = a_f f_{\text{II}}\) (for distribution function of species), \(t_{\text{I}} = a_t t_{\text{II}}\) (for time), \(\mathbf{r}_{\text{I}} = a_r \mathbf{r}_{\text{II}}\) (for position and dimension), \(\mathbf{v}_{\text{I}} = a_v \mathbf{v}_{\text{II}}\) (for velocity), \(q_{\text{I}} = a_q q_{\text{II}}\) (for charge), \(m_{\text{I}} = a_m m_{\text{II}}\) (for mass), \(\mathbf{E}_{\text{I}} = a_E \mathbf{E}_{\text{II}}\) (for electric field), and \(\mathbf{B}_{\text{I}} = a_B \mathbf{B}_{\text{II}}\) (for magnetic field). The scaling coefficients are determined via dimensional analysis in the following.

The Boltzmann equation for system II can be transformed from system I, which is formulated as
% \begin{align}
%     \frac{a_{f_j}}{a_t} \frac{\partial f_j}{\partial t} + &\frac{a_{f_j}a_{v}}{a_r} \mathbf{v} \cdot \nabla f_j + \frac{a_{q}a_{f_j}}{a_m a_v} \frac{q_j}{m_j} \cdot \left( a_E \mathbf{E} + {a_v a_B } {\mathbf{v} \times \mathbf{B}} \right) \cdot \nabla_{\mathbf{v}} f_j \nonumber \\
%     &= (a_{f_j} a_{f_t} a_v^4) \iint (f_{j2} f_{t2} - f_{j1} f_{t1}) \mathbf{v}_{} \sigma_{} \text{d}\Omega \text{d}\mathbf{v}, \label{eq-app-boltII} \\
% \frac{a_E}{a_r} \nabla \cdot \mathbf{E} &= (a_q a_{f_j} a_v^3) \cdot  \frac{1}{\varepsilon_0}\sum_j q_j \int f_j \text{d}\mathbf{v}, \label{eq-app-dotE} \\
% \frac{a_B}{a_r} \nabla \cdot \mathbf{B} &= 0, \\
% \frac{a_E}{a_r} \nabla \times \mathbf{E} &= -\frac{a_B}{a_t} \frac{\partial \mathbf{B}}{\partial t}, \label{eq-app-crossE} \\
% \frac{a_B}{a_r} \nabla \times \mathbf{B} &= {a_q a_{f_j} a_v^4} \mu_0 \sum_j q_j \int \mathbf{v} f_j \text{d}\mathbf{v} + \frac{a_E}{a_t}  \mu_0\epsilon_0 \frac{\partial \mathbf{E}}{\partial t}.
% \label{eq-app-crossB}
% \end{align}
\begin{align}
    \frac{a_{f_j}}{a_t} \frac{\partial f_j}{\partial t} + &\frac{a_{f_j}a_{v}}{a_r} \mathbf{v} \cdot \nabla f_j + \frac{a_{q}a_{f_j}}{a_m a_v} \frac{q_j}{m_j} \cdot \left( a_E \mathbf{E} + {a_v a_B } {\mathbf{v} \times \mathbf{B}} \right) \cdot \nabla_{\mathbf{v}} f_j \nonumber \\
    &= (a_{f_j} a_{f_t} a_v^4) \iint (f_{j2} f_{t2} - f_{j1} f_{t1}) \mathbf{v}_{} \sigma_{} \text{d}\Omega \text{d}\mathbf{v}, 
    \label{eq-app-boltII}
\end{align}
where \((f_{j1}\), \(f_{t1})\) and \((f_{j2}\), \(f_{t2})\) denote the distribution functions of incident and target species before and after collisions.

The full set of Maxwell's equations is transformed by
\begin{align}
\frac{a_E}{a_r} \nabla \cdot \mathbf{E} &= (a_q a_{f_j} a_v^3) \cdot  \frac{1}{\varepsilon_0}\sum_j q_j \int f_j \text{d}\mathbf{v}, \label{eq-app-dotE} \\
\frac{a_B}{a_r} \nabla \cdot \mathbf{B} &= 0, \\
\frac{a_E}{a_r} \nabla \times \mathbf{E} &= -\frac{a_B}{a_t} \frac{\partial \mathbf{B}}{\partial t}, \label{eq-app-crossE} \\
\frac{a_B}{a_r} \nabla \times \mathbf{B} &= {a_q a_{f_j} a_v^4} \mu_0 \sum_j q_j \int \mathbf{v} f_j \text{d}\mathbf{v} + \frac{a_E}{a_t}  \mu_0\epsilon_0 \frac{\partial \mathbf{E}}{\partial t}.
\label{eq-app-crossB}
\end{align}

The Boltzmann and Maxwell equations are scale-invariant if the ratios of the dimensionless parameters are equal in Eq.~\eqref{eq-app-boltII}--\eqref{eq-app-crossB}. From Eq.~\eqref{eq-app-boltII}, the Boltzmann equation is scale-invariant if the dimensionless ratios obey
\begin{equation}
    \frac{a_{f_j}}{a_t} = \frac{a_{f_j} a_v}{a_r} = \frac{a_q a_{f_j} a_E}{a_m a_v}= \frac{a_q a_{f_j} a_v a_B}{a_m}=a_{f_j} a_{f_t} a_v^4.
\label{eq21}
\end{equation}

From Eq.~\eqref{eq-app-dotE}--\eqref{eq-app-crossB}, the full set of Maxwell's equations is scale-invariant if the coefficients follow
\begin{equation}
\frac{a_E}{a_r} = a_q a_{f_j} a_v^3 = \frac{a_B}{a_t}, \quad \frac{a_B}{a_r} = {a_q a_{f_j} a_v^4} = \frac{a_E}{a_t}.
\label{eq22}
\end{equation}

Equations~\eqref{eq21}--\eqref{eq22} are monomials that can be linearized through a logarithmic transformation, which is expressed as
\begin{equation}
\begin{aligned}
\mathbf{Ax} = 0,
\end{aligned}
\label{eq-linearized}
\end{equation}
where \(\mathbf{x}^\text{T}\) = [\(\ln a_r\), \(\ln a_t\), \(\ln a_v\), \(\ln a_q\), \(\ln a_m\), \(\ln a_E\), \(\ln a_B\), \(\ln a_{f_j}\), \(\ln a_{f_t}\)], and \(\text{Rank } \mathbf{ A}=6\) with
\begin{equation}
\begin{aligned}
\mathbf{A}=\left[
\begin{array}{rrrrrrrrr}
1&-1&-1&0&0&0&0&0&0\\
1&0&-2&1&-1&1&0&0&0\\
0&0&-2&0&0&1&-1&0&0\\
0&0&-3&1&-1&0&1&0&-1\\
-1&0&-3&-1&0&1&0&-1&0\\
0&1&3&1&0&0&-1&1&0\\
-1&0&-4&-1&0&0&1&-1&0\\
0&1&4&1&0&-1&0&1&0
\end{array}
 \right].
\end{aligned}
\label{eq-linearized}
\end{equation}
% \begin{scriptsize}
% \begin{equation}
% \begin{aligned}
% \left[
% \begin{array}{rrrrrrrrr}
% 1&-1&-1&0&0&0&0&0&0\\
% 1&0&-2&1&-1&1&0&0&0\\
% 0&0&-2&0&0&1&-1&0&0\\
% 0&0&-3&1&-1&0&1&0&-1\\
% -1&0&-3&-1&0&1&0&-1&0\\
% 0&1&3&1&0&0&-1&1&0\\
% -1&0&-4&-1&0&0&1&-1&0\\
% 0&1&4&1&0&-1&0&1&0
% \end{array}
% \right]\cdot
%  \left[
%  \begin{array}{c}
%  \ln a_r\\ \ln a_t\\ \ln a_v\\ \ln a_q\\ \ln a_m\\ \ln a_E\\ \ln a_B\\ \ln a_{f_j}\\ \ln a_{f_t}
%  \end{array}
%  \right]=\mathbf{Ax} = 0,
% \end{aligned}
% \label{eq-linearized}
% \end{equation}
% \end{scriptsize}
% where \(\text{Rank } \mathbf{ A}=6\) and \(\mathbf{x}^\text{T}\) = [\(\ln a_r\), \(\ln a_t\), \(\ln a_v\), \(\ln a_q\), \(\ln a_m\), \(\ln a_E\), \(\ln a_B\), \(\ln a_{f_j}\), \(\ln a_{f_t}\)].

While eight algebraic relations are derived for nine variables from Eqs.~\eqref{eq-app-boltII}--\eqref{eq-app-crossB}, only six of them are independent.
Among the nine variables, $a_q = 1$ and $a_m = 1$ since the charge and mass of specific species are constant and thus correspondingly identical in the compared systems; $a_r = k$ is predetermined since the geometries of systems I and II are scaled by a ratio of $k$, obeying the similarity requirement.
Plugging in $a_q = 1$, $a_m = 1$, and  $a_r = k$ into Eqs.~\eqref{eq21}--\eqref{eq22} yields six independent equations for six variables, which are expressed as
\begin{equation}
    \frac{k}{a_v a_t}=1, \quad \frac{a_v^2}{ a_E}=k, \quad \frac{a_E}{a_B a_v}=1, \quad \frac{a_E}{a_{f_t} a_v^5}=1,
    \label{eq-app-eqns1}
\end{equation}
\begin{equation}
 \label{eq-app-eqns2}
    \frac{a_E}{a_{f_j} a_v^3}=k, \quad \frac{a_E}{a_{t} a_{f_j} a_v^4}=1,
\end{equation}
where six variables $\{a_v, a_t, a_E, a_B, a_{f_j}, a_{f_t}\}$ are obtained as \(a_v = 1\), \(a_t = k\), \(a_E = k^{-1}\), \(a_B = k^{-1}\), \(a_{f_j} = k^{-2}\), and \(a_{f_t} = k^{-1}\).

According to the similarity transformations, Eqs.~\eqref{eq-G1Gk} and \eqref{eq15}, \(\alpha [r]=1\), \(\alpha [t]=1\), \(\alpha [v]=0\), \(\alpha [q]=0\), \(\alpha [m]=0\), \(\alpha [E]=-1\), \(\alpha [B]=-1\), \(\alpha [f_j]=-2\), and \(\alpha [f_t]=-1\) are determined, which are the similarity factors for the complete set of parameters in weakly ionized plasmas.
However, for strongly ionized plasmas, the Boltzmann equation, when divided by \(k^2\), yields modified similarity factors for the distribution functions of the incident and target species, namely, \(\alpha [f_j]=\alpha [f_t]=-1\), which leave the mathematical solutions invariant.
Consequently, the modified similarity naturally holds as the ionization degree transitions from low to high, given the distinct dominant collisional processes in each regime.

\end{document}